\def   \ni {\noindent}

\def   \ssk {\vskip  5truept}

\def   \bsk {\vskip 15truept}
 
\def   \newpage {\vfill\eject}
\def   \newline {\hfil\break}

\documentstyle[epsfig]{article}
\begin{document}

\hsize 5truein
\vsize 8truein
\font\abstract=cmr8
\font\keywords=cmr8
\font\caption=cmr8
\font\references=cmr8
\font\text=cmr10
\font\affiliation=cmssi10
\font\author=cmss10
\font\mc=cmss8
\font\title=cmssbx10 scaled\magstep2
\font\alcit=cmti7 scaled\magstephalf
\font\alcin=cmr6 
\font\ita=cmti8
\font\mma=cmr8
\def\ref{\par\noindent\hangindent 15pt}
\null


\title{\ni THE INTEGRAL CORE OBSERVING PROGRAMME}

\bsk \bsk
\author{\ni C.~Winkler $^{1}$, N.~Gehrels $^{2}$, N.~Lund $^{3}$,
V.~Sch\"{o}nfelder $^{4}$ and P.~Ubertini $^{5}$}
                                                       
\bsk
\affiliation{\ni (1) ESA/ESTEC, Noordwijk, The Netherlands, 
(2) NASA/GSFC, Greenbelt, USA, (3) DSRI, Copenhagen, Denmark,
(4) MPE, Garching, Germany, (5) IAS, Rome, Italy}
                                                
\bsk
\baselineskip = 12pt

\abstract{ABSTRACT \ni
The Core Programme of the INTEGRAL mission is defined as the portion
of the scientific programme covering the guaranteed time observations
for the INTEGRAL Science Working Team. This paper describes
the current status of the Core Programme preparations 
and summarizes the key elements of the observing programme.
}                                                    

\bsk
\baselineskip = 12pt

\bsk
\baselineskip = 12pt


\text{\ni 1. INTRODUCTION
\ssk
\ni     

Scientific observing time for the observing programme during 
the nominal (2 yr) and extended (3 yr) mission phases of INTEGRAL is 
divided into the open time for the General Observer
(General Programme; year 1: 65\%, year 2: 70\%, year 3+: 75\%) 
and the guaranteed time for the INTEGRAL Science Working Team (ISWT)
(Core Programme; year 1: 35\%, year 2: 30\%, year 3+: 25\%).
Observing time for the General Programme will be allocated to the 
scientific community during the AO process.
The Core Programme will consist of three elements
(see also Gehrels et al. 1997, Ubertini et al. 1997a):
(i) Frequent scans of the Galactic plane (Galactic Plane Survey, GPS)
(ii) Deep exposure of the Galactic central radian 
(Galactic Central Radian Deep Exposure, GCDE)
(iii)) Pointed observations of selected sources. 
The data from the Core Programme 
belong to the ISWT for the usual proprietary period of one 
year, after which the data will become public.

\bsk
\ni 2. SCIENCE OBJECTIVES OF THE CORE PROGRAMME
\ssk
\ni

The \underline{scanning of the Galactic plane (GPS)} will 
be mainly done for two reasons: 
the most important one is to provide frequent monitoring of the plane in 
order to detect transient sources because the gamma-ray sky in the INTEGRAL 
energy range is dominated by the extreme variability of many sources. The 
scans would find sources in high state (outburst) which warrant possible 
scientifically important follow-up observations (Target of Opportunity [TOO] 
observation). The second reason is to build up 
time resolved maps of the Galactic plane in 
continuum and diffuse line emission such as $^{26}$Al and 511 keV with modest 
exposure. 
Compton GRO and SIGMA have been detecting Galactic compact sources of 
several different categories/groups which include 
X-ray binaries (e.g. X-ray novae, Be binary pulsars) and in particular
superluminal sources (GRS 1915$+$105 and GRO J1655$-$40). 
The occurrence rates for events that INTEGRAL can observe 
is about 2 events/year for 
\newpage
\ni each of these classes, where pointing constraints 
due to the fixed solar arrays have been taken into account. The important 
time scales for the transient outbursts vary significantly from class to 
class and from event to event, but a typical duration of an event is 1 - 2 
weeks and a typical variability time scale is of the order of 1 day. 
The scans will be 
performed once a week by performing a ``slew - and stare'' manoeuvre 
of the spacecraft along the visible (accessible) part of the 
Galactic plane with latitude extent $\pm10^{\circ}$.
The accessible part of the Galactic plane depends on viewing constraints 
including the solar aspect angle (40$^{\circ}$ during first two years,
30$^{\circ}$ thereafter) due to the fixed solar arrays, 
and on the season of the year. 
The angular distance 
between subsequent exposures (965 s each) along the scan path is 6$^{\circ}$. 
The scan will be performed as a sawtooth 
with inclination of 21$^{\circ}$ with respect to the Galactic plane, 
each subsequent scan is shifted by 27.5$^{\circ}$ in galactic longitude.

The 
\underline{deep exposure of the central Galactic radian (GCDE)} 
is driven by the following 
objectives: mapping line emission from nucleosynthesis radioisotopes 
(e.g. $^{26}$Al, $^{44}$Ti, 511 keV), mapping continuum emission of the Galactic 
ridge, and performing deep imaging and spectroscopic studies of the 
central region of the Galaxy. Several interesting emission regions 
in or near the Galactic plane have been identified using CGRO OSSE and 
COMPTEL: these include the $^{26}$Al (7$\times$10$^5$ years half life) mapping the 
sites of nucleosynthesis in the past million years in the Galaxy 
(Diehl et al. 1995), and $^{44}$Ti (half life $\sim$60 year) 
which has been detected by COMPTEL from the Cas A SNR 
(Iyudin et al. 1994). OSSE mapping of the 
positron - electron annihilation radiation at 511 keV shows a 
central bulge, emission in the Galactic plane and an enhancement of 
extension of emission at positive latitudes above the Galactic 
centre (Purcell et al. 1997). Other isotopes such as $^{60}$Fe 
produce lines which could be detected by INTEGRAL. 
The origin of the clumpy structure of the COMPTEL observed 
$^{26}$Al maps and the $^{44}$Ti emission from hidden supernovae are 
key targets of INTEGRAL research.
The INTEGRAL deep exposure will also study the continuum gamma-ray 
and hard X-ray emission from the Galactic plane. This ``galactic ridge'' 
is concentrated in a narrow band with a latitude extent of $\sim5^{\circ}$ 
and a longitude extent of $\pm40^{\circ}$ (Gehrels \& Tueller 1993, 
Valinia \& Marshall 1998, Strong et al. 1999).
The exact distribution and spectrum of the ridge emission is not 
well known. The origin is thought to be Bremsstrahlung from cosmic 
ray electrons, but this is also not fully established. INTEGRAL 
will be able to map the emission with high sensitivity and high 
angular resolution. This should allow the removal of the point-source 
origin of the emission so that the spectrum can be determined              
with high confidence. 
The GCDE will resolve isolated point sources with arcmin location and 
provide source spectra with high energy resolution. At least 90 
sources known as X- and gamma-ray emitters are contained in the 
region at $\sim10$ keV and GRANAT/SIGMA and other earlier experiments 
have found this region to be filled with a number of highly 
variable and transient sources (Vargas et al. 1997,
Ubertini et al. 1997).
Many are thought to be compact objects in binary systems 
undergoing dynamic accretion. However, even for some of the 
brightest sources in the region (1E 1740.7$-$2942, GRS 1758$-$258) 
the detailed nature of the systems are not known. 
INTEGRAL will study the faintest sources with 
high angular resolution allowing multi-source monitoring 
within its wide field of view during single pointings. 
Also interesting will be searches for gamma-ray emission from 
SgrA$^*$, at the Centre of the Galaxy 
(Sunyaev et al. 1993).
The central radian of the Galaxy will be observed using a rectangular 
pointing grid of 11 $\times$ 31 = 341 pointings with a pitch of 2$^{\circ}$. 
The grid covers the celestial region between -30$^{\circ}$ $\le$ l $\le$ 
+30$^{\circ}$ and -10$^{\circ}$ $\le$ b $\le$ +10$^{\circ}$. 
With 3 minutes slew duration for 2$^{\circ}$ and 1000 s 
exposure per point, about 4$\times$10$^5$ s will be required to 
scan the grid once. 
Within the total time allocation for the GCDE, 
the grid scan will be performed 12 times per year. 
Remaining Core Programme observing time not spent on GPS and GCDE will 
be devoted to \underline{dedicated pointings on individual sources}. 
These sources 
include galactic black hole candidates, neutron stars, AGN
and some time will 
also be set aside to be able to perform scientifically important 
follow-up observations on TOO's which have been detected during 
GPS and GCDE or where the trigger has been observed by optical telescopes 
(SN and Novae).  
The greatest advantage of INTEGRAL, as compared to other high 
energy missions, is its high spectral resolution and 
high sensitivity at energies above 100 keV. 
The strongest point for 
INTEGRAL to response to a TOO call is the potential to detect 
transient high energy emission features (e.g. 511 keV, its scattering feature 
at 170 keV; 480 keV ($^7$Li); MeV bump; flares from Blazars).
The high energy cutoff or spectral break as a function of time may 
provide critical information to our understanding of the emission 
mechanism and system parameters. 
Concerning the high energy continuum, several pointings for each flare or 
outburst are 
required to achieve meaningful science return since the 
spectral evolution instead of simply the spectral shape is the 
main objective here. Separation between the exposures depends on 
the time scale of the event.
The final list of candidate sources 
will be published, together with 
all details of the other Core Programme elements, 
in the first AO for INTEGRAL observations, which is scheduled for release 
at the end of 1999.
General Observers cannot propose for targets inlcuded in the
Core Programme list of candidate sources, unless the scientific objectives 
of their proposals are of a clearly different nature.


The orbit characteristics for INTEGRAL on PROTON (baseline) and ARIANE 5 
(backup) are summarised in Winkler (1999) and Carli (1999). 
The allocation of observing time
has been driven by highest scientific 
priority which has been given to GPS and GCDE and is presently scheduled
for the first year of operations (baseline orbit) as follows:
GCDE: 4.8$\times$10$^6$~s; GPS: 2.3$\times$10$^6$~s; Pointed 
observations: 2.2$\times$10$^6$~s assuming an observation efficieny of 85\%.
It is planned to keep the allocations constant for GCDE and GPS
during the year 2 of operations (and for GCDE during year 3 to 5).



\bsk
\ni 3. EXPOSURE TIMES AND SENSITIVITIES
\ssk
\ni

Due to the large coded field-of-views for SPI (FCFOV 16$^{\circ}$ across)
and IBIS (FCFOV 9$^{\circ}$ $\times$ 9$^{\circ}$), a single ``staring'' 
point on a \underline{GPS scan} will not only receive the nominal exposure per 
pointing (965 s):
While this exposure point is 
still located within the partially coded 
field-of-view (PCFOV, coverage to zero response), 
additional exposure will be collected as the spacecraft 
(i.e. the co-aligned instruments) is pointing at close neighbouring 
positions of that single exposure point.

For SPI and IBIS, a single ``staring'' point will therefore 
receive $\sim$3000 s of exposure per scan, while JEM-X and OMC -- because
of the smaller FOV's -- will accumulate 965 s per point/scan.
The 3$\sigma$ sensitivity which can be obtained 
for one ``staring'' point during a single scan is 
therefore 10 mCrab @ 100 keV (IBIS), 
750 mCrab @ 1 MeV (SPI) and 6 mCrab @ 10 keV (JEM-X). 
This will allow detection of transient sources 
throughout the Galaxy.
During the first two years the total exposure per point per year
is $\sim$9$\times$10$^{4}$ s for SPI and IBIS. 
For JEM-X and OMC the exposure per point
and year amounts to 2.7$\times$10$^{4}$ s.
The area covered by the \underline{GCDE} 
will receive a net exposure of 4$\times$10$^6$ s per year 
which will give SPI a 3$\sigma$ sensitivity of 
$\sim$5$\times$10$^{-6}$ ph/(cm$^2$s) for narrow 
lines in the 100 keV - 2 MeV region sufficient for mapping and for 
detailed line shape studies at 1809 keV (Gehrels \& Chen 1996)
of the bright $^{26}$Al ``hot spots'' ($\sim$3$\times$10$^{-5}$ ph/(cm$^2$ s)) 
as detected by COMPTEL.
The continuum sensitivities (3$\sigma$) in the 100 keV to 1 MeV range would 
be 
5$\times$10$^{-7}$ to 4$\times$10$^{-8}$ ph/(cm$^2$ s keV) (SPI) and 
(2 to 1)$\times$10$^{-7}$ ph/(cm$^2$ s keV)
(IBIS) while JEM-X would achieve 10$^{-5}$ to 10$^{-6}$ 
ph/(cm$^2$ s keV) in the 3 to 30 keV range.
\underline{Pointed observations}, 
e.g. nucleosynthesis studies of Cygnus and Vela regions require 
typically $\sim$10$^6$ s while Galactic compact (transient) objects 
could be observed with 10$^5$ s or less per repeated observation. 
Given the high sensitivity of INTEGRAL during the 
GPS (and GCDE), follow-up observations for Galactic TOO's - like 
the superluminal jet source GRS 1915$+$105  or Cyg X$-$1 - could be 
triggered if the 100 keV flux would exceed a few hundred mCrab, 
easily detectable in one GPS exposure.
Supernovae and Novae observations would be possibly done repeatedly
at various times after onset, to study (e.g. for SNII) early observations 
of radioactivity from Co breaking through the SN envelope and 
late observations of $^{44}$Ti.
}

\baselineskip = 12pt
{\abstract \ni ACKNOWLEDGMENTS
\ni We thank G.~Skinner and P.~Connell for simulation studies.
}

\baselineskip = 12pt


{\references \ni REFERENCES
\ssk


\ref Carli R et al., 1999, these proceedings
\ref Diehl R., et al., 1995, A\&A 298, 445
\ref Gehrels N., Tueller J., 1993, ApJ 407, 597
\ref Gehrels N., Chen W., 1996, A\&AS 120, 331
\ref Gehrels N., et al., 1997, ESA SP-382, 587
\ref Iyudin A., et al., 1994, A\&A 284, L1
\ref Purcell W., et al., 1997, ApJ 491, 725
\ref Strong A., et al., 1999, these proceedings
\ref Sunyaev et al., 1993, ApJ 407, 606.
\ref Ubertini P., et al., 1997, Proc. 4th CGRO Symposium, AIP 410, 1527
\ref Ubertini P., et al., 1997a, Proc. Workshop on 'ASM and GRB emission'
\ref Valinia A., Marshall F., 1998, ApJ 505, 134
\ref Vargas M., et al., 1997, ESA SP-382, 129
\ref Winkler C., 1999, these proceedings
}
\end{document}